# THEORETICAL ASSESSMENT OF THE DISPARITY IN THE ELECTROSTATIC FORCES BETWEEN TWO POINT CHARGES AND TWO CONDUCTIVE SPHERES OF EQUAL RADII

*Kiril Kolikov*
Plovdiv University "Paisii Hilendarski", Plovdiv, Bulgaria

**Abstract**

The Coulomb's formula for the force $F_C$ of electrostatic interaction between two point charges is well known. In reality, however, interactions occur not between point charges, but between charged bodies of certain geometric form, size and physical structure. This leads to deviation of the estimated force $F_C$ from the real force $F$ of electrostatic interaction, thus imposing the task to evaluate the disparity.

In the present paper the problem is being solved theoretically for two charged conductive spheres of equal radii and arbitrary electric charges. Assessment of the deviation is given as a function of the ratio of the distance $R$ between the spheres centers to the sum of their radii. For the purpose, relations between $F_C$ and $F$ derived in a preceding work of ours, are employed to generalize the Coulomb's interactions.

At relatively short distances between the spheres, the Coulomb force $F_C$, as estimated to be induced by charges situated at the centers of the spheres, differ significantly from the real force $F$ of interaction between the spheres.

In the case of zero and non-zero charge we prove that with increasing the distance between the two spheres, the force $F$ decrease rapidly, virtually to zero values, i.e. it appears to be short-acting force.

## 1. Introduction

The formula for the magnitude of Coulomb force $F_C = \dfrac{Q_1 Q_2}{4\pi\varepsilon_0 R^2}$ of electrostatic interaction in vacuum between two point charges $Q_1$ and $Q_2$ at a distance $R$ from each other are well known [1], [2]. In reality, interactions take place not between point charges, but between charged bodies of certain geometric form, size and physical structure. It was Maxwell who discovered that electrostatic force between two spheres differed from the electrostatic force between point charges of the same magnitude located at the centres of the spheres [3, Chapter 1]. In his opinion it resulted from the redistribution of charges, due to the mutual electrostatic influence of the spheres. In this way the problem of assessing the deviation of $F_C$ from the actual values of the force $F$ of electrostatic interaction arises, especially at relatively short distances.

Using the capacitance coefficients, Lekner [4], [5], [6] investigate the case of spheres at close-approach and derive expression for the force between two spheres. At that, using formula (3.4) from [5], he derives the leading term of the force $F$. If $r_1 = r_2 = r$ and $Q_1 \neq Q_2$ this term is

(1)
$$F_L = -\frac{1}{4\pi\varepsilon_0} \frac{(Q_1 - Q_2)^2}{2r(R - 2r)\left[\ln\left(\frac{4r}{R - 2r}\right) + 2\gamma\right]^2}.$$

Here $\gamma = 0.577...$ is Euler-Mascheroni's constant. Those expressions generalize the force formula for obtained by Kelvin [7].

Khair [8] derive analytical expressions for the electrostatic forces on two almost touching nonspherical conductors held at unequal voltages or carrying dissimilar charges in an insulating medium. Each of those conductors is a body of revolution whose surface is defined by the equation $r^n + z^n = a^n$, where $r$ and $z$ are radial and axial cylindrical co-ordinates, respectively, and $a$ and $n$ are parameters that control the particle width and shape.

In the particular case of spheres with $r_1 = r_2 = r$ and $Q_1 \neq Q_2$ for the leading term of the force $F$ he derives formula (17) from [8]

(2)
$$F_K = -\frac{(Q_1 - Q_2)^2}{16\varepsilon_0 \pi r^2}\left(\frac{2r}{R - 2r}\right)\frac{1}{\left[\ln\left(\frac{2r}{R - 2r}\right)\right]^2}.$$

Khair notes that his formula is not as accurate as Lekner's.

From formulas (1) and (2) follows that in a close enough distance between the surfaces of the spheres, force $F$ is always a force of attraction.

Using the method of image charges in [9], we were the first to express by infinite sums the *exact analytical formula* for the force $F$ of electrostatic interaction between two conductive spheres of arbitrary charges $Q_1$, $Q_2$ and radii $r_1$, $r_2$ (as well as the energy $W$ and the potential $V$ at a given point of the field induced by them). From this general formula for $r_1 = r_2 = 0$, follows the Coulomb's interaction!

In this paper the problem of determining the divergence in the magnitude of the Coulomb force $F_C$ from the actual values $F$ of electric interaction between two charged conductive spheres of arbitrary charges $Q_1$, $Q_2$ and equal non-zero radii ($r_1 = r_2 = r$) is being solved. It is assumed that the force $F_C$ is generated by $Q_1$ and $Q_2$ located in the center of the sphere.

With non-zero charges $Q_1$ and $Q_2$ of the spheres in [9], [10] we have shown already that $F = F_C L$, where $L$ is dimensionless coefficient. In [11] we have mathematically analyzed the convergence of the coefficient of deviation $L$.

Here using the formula for $L$ from [9], [10] the possible values of the deviations of $F_C$ from $F$ are investigated by means of graphical analysis for different non-zero charges $Q_1$ and $Q_2$ as a function of the ratio of the distance $R$ between the spheres to the sum $2r$ of their radii.

The conclusion of formulas (1) and (2) is refined, as with given like charges $Q_1 \neq Q_2$, the accurate values of $R/2r$ in which $F$ becomes force of attraction are found.

By introducing another type dimensionless coefficient $L(0)$ the deviation of $F_C = 0$ from $F$ is studied also in the case when one of the charges is zero. It is proven in this case that with distance increase between spheres $F$ rapidly decrease virtually to zero values, i.e. it operates at comparatively short distances.

## 2. Relation between the Coulomb's force and the electrostatic force between two conductive spheres of equal radii

Let $S_1$ and $S_2$ are two non-intersecting conductive spheres of equal non-zero radii $r_1 = r_2 = r$ and arbitrary charges $Q_1$ and $Q_2$.

If $R$ is the distance between centers $O_1$ and $O_2$ of the spheres $S_1$ and $S_2$ in inertial frame of reference $J$, then in [9] we put $\delta = \dfrac{r}{R}$ and denote:

$$(3) \quad C_j = \frac{\left(1+\sqrt{1-(2\delta)^2}\right)^{j+1} - \left(1-\sqrt{1-(2\delta)^2}\right)^{j+1}}{2^{j+1}\sqrt{1-(2\delta)^2}}, \quad j = 0,1,2,\ldots$$

Also:

$$(4) \quad X = \sum_{m=1}^{\infty} \frac{\delta^{2m}}{C_{2m}}, \quad Y = \sum_{m=1}^{\infty} \frac{\delta^{2m-1}}{C_{2m-1}}.$$

If the charges of spheres $S_1$ and $S_2$ are respectively $Q_1 \neq 0$ and $Q_2$, we can introduce a coefficient $k = \dfrac{Q_2}{Q_1}$, representing the ratio of the charges.

**2.1.** Let $Q_1 \neq 0$ and $Q_2 \neq 0$, e.g. $k \neq 0$.

Then from [9] we find that the image charges are situated at distances $d'_j$ and $d''_j$ relative to the centers of spheres $S_1$ and $S_2$, where $d'_j = d''_j = d_j = \delta^2 R \dfrac{C_{j-1}}{C_j}$, $j = 0,1,2,\ldots$

We put:

$$(5) \quad L'_0 = \frac{1+X+kY}{(1+X)^2 - Y^2}, \quad L''_0 = \frac{1+X+k^{-1}Y}{(1+X)^2 - Y^2}$$

and for $m = 1,2,3\ldots$

$$(6) \quad L'_{2m-1} = -\frac{\delta^{2m-1}}{C_{2m-1}}kL''_0, \quad L'_{2m} = \frac{\delta^{2m}}{C_{2m}}L'_0, \quad L''_{2m-1} = -\frac{\delta^{2m-1}}{C_{2m-1}}k^{-1}L'_0, \quad L''_{2m} = \frac{\delta^{2m}}{C_{2m}}L''_0.$$

Let $\delta_j = \dfrac{d_j}{R} = \delta^2 \dfrac{C_{j-1}}{C_j}$, $j = 0,1,2,\ldots$ Then from [9], [10] we obtain that force $F$ of electrostatic interaction between two spheres is represented in the form

$$(7) \quad F = F_C L(k).$$

Here $F_C$ is determined for point charges $Q_1$ and $Q_2$ located respectively in the centers $O_1$ and $O_2$ of the spheres, and

$$(8) \qquad L(k) = \sum_{j=0}^{\infty} \sum_{i=0}^{\infty} \frac{L'_j L''_i}{\left(1 - \delta_j - \delta_i\right)^2} \quad (k \neq 0)$$

is dimensionless coefficient of proportionality between $F$ and $F_C$.

*The coefficients of deviation* $L(k)$ is expressed by two parameters: the ratio of the charges $k = \frac{Q_2}{Q_1} \neq 0$, and the ratio $\delta = \frac{r}{R}$ of the spheres radii to the distance $R$ between their centers.

**2.2.** Let $Q_1 = Q \neq 0$ and $Q_2 = 0$, e.g. $k = 0$.

Then deviation coefficients of $F_C = 0$ from $F$ do not exist. In this case in [9] we have shown that, relative to spheres' centers $S_1$ and $S_2$, the image charges are situated at distances $d'_j$ and $d''_j$, where:

$$(9) \qquad d'_0 = d''_0 = 0, \quad d'_{2m} = \delta^2 R \frac{C_{2m-1}}{C_{2m}}, \quad d''_{2m-1} = \delta^2 R \frac{C_{2m-2}}{C_{2m-1}}, \quad d'_{2m-1} = d''_{2m} = 0,$$

e.g. $\delta'_j = \frac{d'_j}{R} \neq \delta''_j = \frac{d''_j}{R}$, $j = 1, 2, 3, ...$

Furthermore:

$$(10) \qquad L'_0 = \frac{1}{1+X}, \quad L''_0 = \frac{Y}{1+X}, \quad L'_{2m} = \frac{\delta^{2m}}{C_{2m}} L'_0, \quad L''_{2m-1} = -\frac{\delta^{2m-1}}{C_{2m-1}} L'_0, \quad L'_{2m-1} = L''_{2m} = 0.$$

From these equations, in accord with formulas (7), (9) and (10) from [9], we can derive for the force $F$:

$$(11) \qquad F = F'_C L(0),$$

where

$$(12) \qquad F'_C = \frac{Q^2}{4\pi\varepsilon_0 R^2},$$

and

$$(13) \qquad L(0) = \sum_{j=0}^{\infty} \sum_{i=0}^{\infty} \frac{L'_j L''_i}{\left(1 - \delta'_j - \delta''_i\right)^2}$$

is dimensionless coefficient.

*The supplementary coefficient* $L(0)$ is only expressed through one parameter: the ratio $\delta = \frac{r}{R}$ and define deviation of $F'_C$ from $F$. Hence, $L(0)$ can be used for juxtaposing $F_C$ and $F$, taking into account the deviation of $F'_C$ from $F_C = 0$.

Lets note that with point charges $r_1 = r_2 = 0$ i.e. $\delta_1 = \delta_2 = 0$. It is only in this case, when from formulas (4), (5), (8) and formulas (10) and (13), it follows that $L(k) = 1$ ($k \neq 0$) and $L(0) = 0$ ($k = 0$). Then, according to the formulas (7) and (11) $F = F_C$.

Therefore, equations (7) and (11) appear to be *generalizations of Coulomb forces*!

**3. Deviation of the Coulomb's force from the force of interaction between two spheres with equal radii.**

We will determine the deviation of the Coulomb's force $F_C$ and the electrostatic force $F$ between two conducting spheres $S_1$ and $S_2$ with equal radii $r_1 = r_2 = r$, charged with charges $Q_1$ and $Q_2$. To this end we will discuss the possible values, which it takes, according to the formulas (8) and (13), the respective deviation coefficient $L(k)$ with different values of the real numbers $k = Q_2/Q_1$ ($Q_1 \neq 0$).

We will study the non-dimensional coefficient $L(k)$ depending on the ration of the distance $R$ between the centers of the spheres $S_1$ and $S_2$ and the sum $2r$ of their radii, i.e. depending on $(2\delta)^{-1}$. (By $2\delta$, $C_j$ is calculated from formula (3), and hence $L(k)$ from formulas (8) and (13)). It is known, that for nonintersecting spheres $\frac{R}{2r} > 1$, i.e. the distance between the surfaces of the spheres $R - 2r > 0$.

Since the spheres' radii are equal, it is sufficient to discuss $L(k)$ only with $-1 \leq k \leq 1$, i.e. when $|Q_1| \geq |Q_2|$.

We will discuss the two main cases of charged spheres - with unlike and like charges. According to formulas (8) and (13) on Fig. 1, we present the areas for $L(k)$ with $-1 \leq k < 0$ and $0 \leq k \leq 1$, depending on $R/2r$.

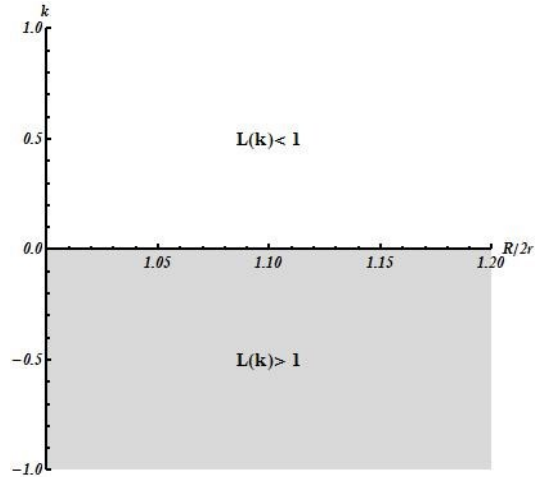

Fig. 1. Areas for $L(k)$, defined by $R/2r$, with $-1 \leq k < 0$ and $0 \leq k \leq 1$.

It could be seen that $L(k) > 1$ for each $-1 \leq k < 0$ (in the fourth quadrant) and $L(k) < 1$ for each $0 \leq k \leq 1$ (in the first quadrant). From here and from the formulas (7) and (11), it follows that $F_C \neq F$ for arbitrary values of $k = Q_2/Q_1$ and $R/2r$.

From $L(k) > 1$ follows that for unlike charges $F < F_C < 0$, and then $|F_C| < |F|$. Therefore,

for *unlike charges* the force of attraction between the spheres is greater than the Coulomb force.

We discuss the case $L(k)<1$ in more detail, i.e. regarding spheres charged with different *like charges*.

According to formula (8), we present on Fig. 2 the areas of $L(k)$ for $0<k<1$, defined by $R/2r$.

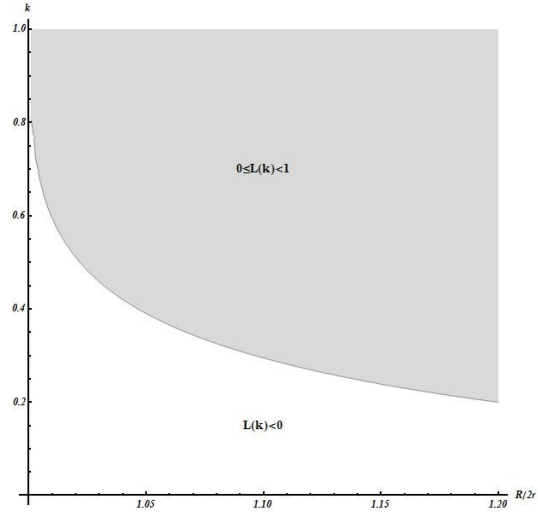

Fig. 2. Areas of $L(k)$ for $0<k<1$, defined by $R/2r$.

Fig. 2 shows that with different like charges $Q_1$ and $Q_2$, the coefficient $L(k)$ and hence the force $F$ change their sign, i.e. $F \geq 0$ or $F<0$. By increasing the ratio $Q_2/Q_1 <1$, the values of $R/2r>1$ with which $L(k)<0$ (and hence $F<0$) decrease. Therefore, for small enough distances $R-2r>0$ between the surfaces of the spheres $F<0$, while $F_C>0$ for each $k>0$.

For example, $F<0$ with: $k<0.59$ and $R/2r<1.01$; $k<0.804$ and $R/2r<1.001$, $k<0.9169$ and $R/2r<1.0001$, $k<0.96741$ and $R/2r<1.00001$, etc.

This agrees with formula (3.4) from [5], that upon contact of the spheres the force $F<0$. We specify this result as with formula (8) we can find the exact values of $R/2r$ at which for given positive $k=Q_2/Q_1$, $F$ becomes force of attraction. (Of course, at very close distances between the spheres practically there will be a spark, and there will be no attraction).

We will discuss the cases $k=0,\pm 1$ of the represented areas of $L(k)$.

Along with $k=0,\pm 1$, the graphics of the coefficients $L(k)$ with $k=\pm\dfrac{1}{4},\pm\dfrac{1}{16}$, depending on $R/2r$, are shown for comparison on Fig. 3.

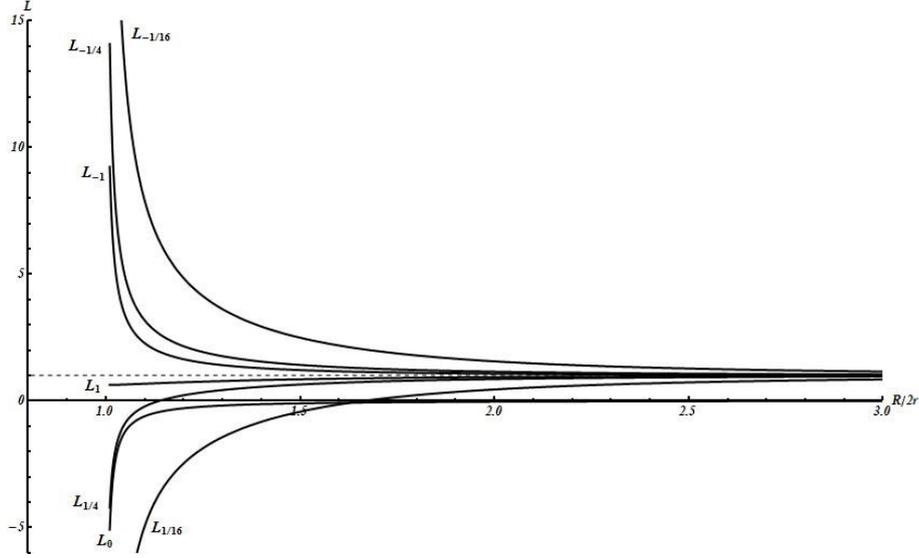

Fig. 3 Graphics of the coefficients $L(k)$, with different $k = Q_2/Q_1$, depending on $R/2r$.

It is visible that the values of $L(-1)$ and $L(1)$ are the smallest respectively for $k<0$ and $k>0$, and $L(-1) > L(1)$ for each $R/2r$. Therefore $|F(-1)| > F(1)$ while for the Coulomb force $|F_C(-1)| = F_C(1)$ is valid.

Since it is calculated that $L(1) \in (0.6;1)$ and the deviation of $F_C$ from $F$ is the smallest of all values of $k$, then only with $Q_1 = Q_2$ it could be assumed with certain accuracy that $F_C = F$ for each $R/2r$.

When $k=0$, i.e. $Q_2 = 0$ it is visible that $L(0) < 0$ with each value of $R/2r$. Hence $F(0) < 0$ while $F_C = 0$ and then $F_C < |F(0)|$. Thus, $F(0)$ is a short-acting force and it is growing very fast when $R/2r \to 1$.

Moreover, the graphs on Fig. 3 show that at close distances, as a result from the values of $L(k)$, $F_C$ significantly deviates from $F$, as the deviation is bigger for unlike charges.

It is important for the practice to determine when, with a given accuracy, it could be assumed that $F_C = F$. Therefore, based on formula (8), we will determine the values of $k \neq 0$, depending on $R/2r$ which, with the accuracy to the second decimal place, the coefficient $L(k) = 1$.

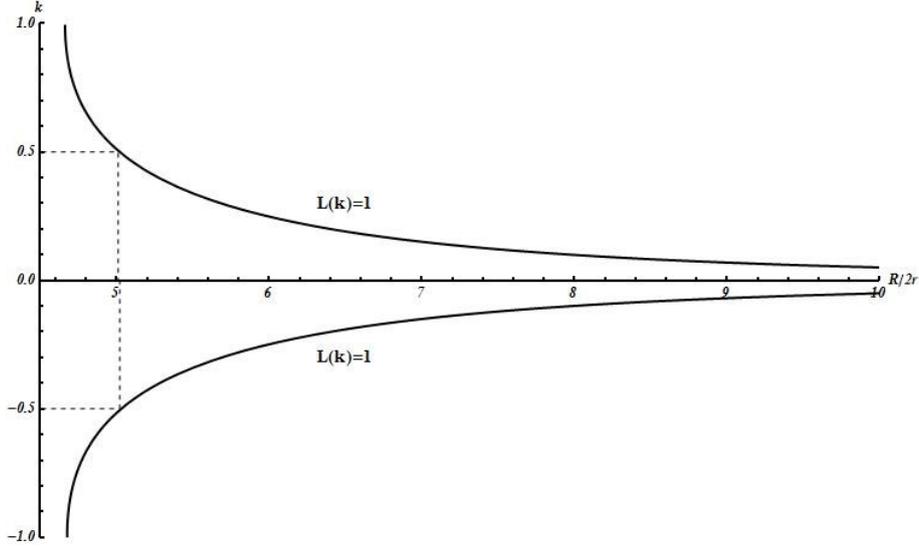

Fig. 4. Graphics of the relation $k = Q_2/Q_1$ depending on $R/2r$, at $L(k)=1$.

From the graphics of Fig. 4, it follows that at $R/2r > 4.65$, i.e. for relatively big values of $R$ compared $2r$, with accuracy to the second decimal place $L(k)=1$, i.e. $F_C = F$. And at $R/2r \leq 4.65$, $F_C \neq F$ with accuracy to the second decimal place. We can see also that when $k<0$ and $k>0$ the two graphics are almost symmetric. For example, when $k = \pm 1/2$ and $R/2r \approx 5$ the coefficient $L(k)=1$, i.e. $F_C = F$, with accuracy to the second decimal place.

Moreover, it is apparent that with increase in modulus of the relation $k \in [-1,1]$ between the charges, there is a decrease of the values of $R$ in relation to $2r$ in the interval $(4.65; +\infty)$, at which $F_C \approx F$. Therefore with $k \to 0$ the deviation of $F_C$ from $F$ is the biggest, as follows from Fig. 3.

Of course, if the accuracy of the study is increased, then $L(k)=1$ in much bigger values of $R/2r$. But we can always determine it for two spheres, with given equal radii $r$ and charges $Q_1$ and $Q_2$, at what distances $R$ compared $2r$ the electrostatic force $F$ can be assumed to be equal to the Coulomb's force $F_C$, caused by these charges in the centers of the spheres.

**4. Conclusions and Example**
Let $S_1$ and $S_2$ be two conducting spheres with equal radii $r_1 = r_2 = r \neq 0$ and arbitrary charges $Q_1$ and $Q_2$. Then we can assume, that $k = Q_2/Q_1 \in [-1,1]$ and according to Fig. 1-4, are valid the following important conclusions regarding the deviation of the Coulomb's force $F_C$ from the real force $F$, when the distance between $S_1$ and $S_2$ is changing.

**Conclusions regarding the deviation of $F_C$ from *F*:**
*(i)* When the values of $Q_1$ and $Q_2$ are arbitrary, $F_C \neq F$ holds, as for $R/2r \to 1$ the deviation of $F_C$ from $F$ grows rapidly.

*(ii)* With $Q_1 \neq 0$ and $Q_2 \neq 0$:

- For like charges $Q_1 \neq Q_2$, i.e. for $F_C > 0$, the force $F$ changes its sign as with $R/2r \rightarrow 1$ the force $F < 0$ and $F_C < |F|$ and for relatively bigger values of $R/2r$ the force $F \geq 0$ and $F_C > F$.
- For unlike charges $Q_1$ and $Q_2$, i.e. for $F_C < 0$, the force $F < 0$ and $0 < |F_C| < |F|$. In this case the absolute value of the deviation is bigger than the one for like charges.
- For $Q_1 \rightarrow \pm Q_2$, decrease the values of $R/2r$, at which $F_C \approx F$, as for $Q_1 = Q_2$ the force $F > 0$ and the deviation of $F_C$ from $F$ is the smallest.

*(iii)* With $Q_1 \neq 0$ and $Q_2 = 0$, i.e. for $F_C = 0$, the force $F < 0$ is short-acting, and then $|F_C| < |F|$.

In the next example we graphically compare the values of $F_C$ with the values of $F$, $F_L$ and $F_K$, respectively of formulas (7), (1) and (2), with various non-zero like charges.

**Example 1.** Graphics of the forces $F_C$, $F$, $F_L$ and $F_K$ for $r_1 = r_2 = 2 \times 10^{-2}$ m and $Q_1 = 16 \times 10^{-9}$ C, $Q_2 = 4 \times 10^{-9}$ C, for $R \in (4,5] \times 10^{-2}$ m.

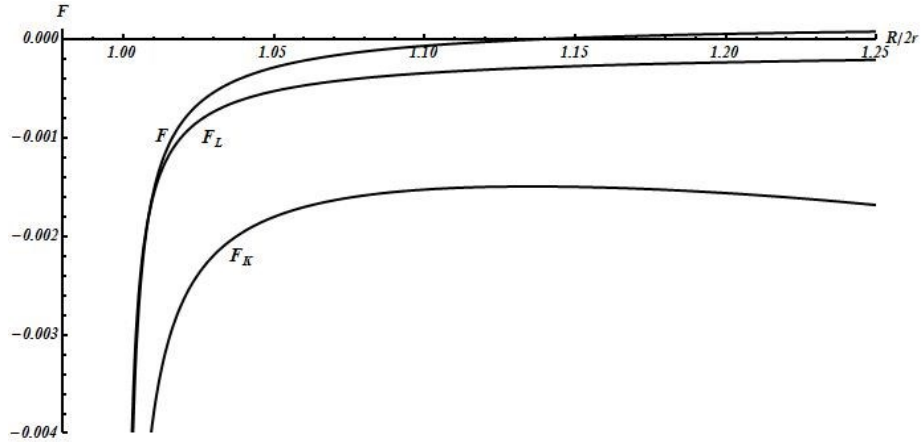

Fig. 5 Graphics of $F_C$, $F$, $F_L$ and $F_K$, depending on $R/2r$, for spheres with radii $r_1 = r_2 = 2 \times 10^{-2}$ m and charges $Q_1 = 16 \times 10^{-9}$ C, $Q_2 = 4 \times 10^{-9}$ C, for $R \in (4,5] \times 10^{-2}$ m

We can see, that $F_C$ substantially deviates from the values of $F$, $F_L$ and $F_K$. For relatively close distance $R = 4.05 \times 10^{-2}$ m between the centers of the spheres, i.e. for $R/2r = 1.0125$, we have $F/F_C = -3.62$, $F_L/F_C = -4.74$ and $F_K/F_C = -9.61$.

The deviation between the spheres is smallest when using our exact analytical formula (7) for $F$. The differences between the obtained values are due to the inaccuracy of the used methods for $F_L$ and $F_K$. But when the distances are close enough, the values of $F$ from formula (7) and $F_L$ from the derived for that purpose formula (1) match.

In the next example about $Q_1 = Q_2 = Q$ it is not possible to compare the values of $F/F_C$, $F_L/F_C$ and $F_K/F_C$, since the formulas (1) for $F_L$ and (2) for $F_K$ are valid only when $Q_1 \neq Q_2$. Moreover, $F_C = F'_C = \dfrac{Q^2}{4\pi\varepsilon_0 R^2}$ from the formula (12).

**Example 2.** The graphics of the forces $F_C$ and $F$ with $r_1 = r_2 = 2\times 10^{-2}$ m and $Q_1 = Q_2 = 16\times 10^{-9}$ C for $R \in (4,5]\times 10^{-2}$ m.

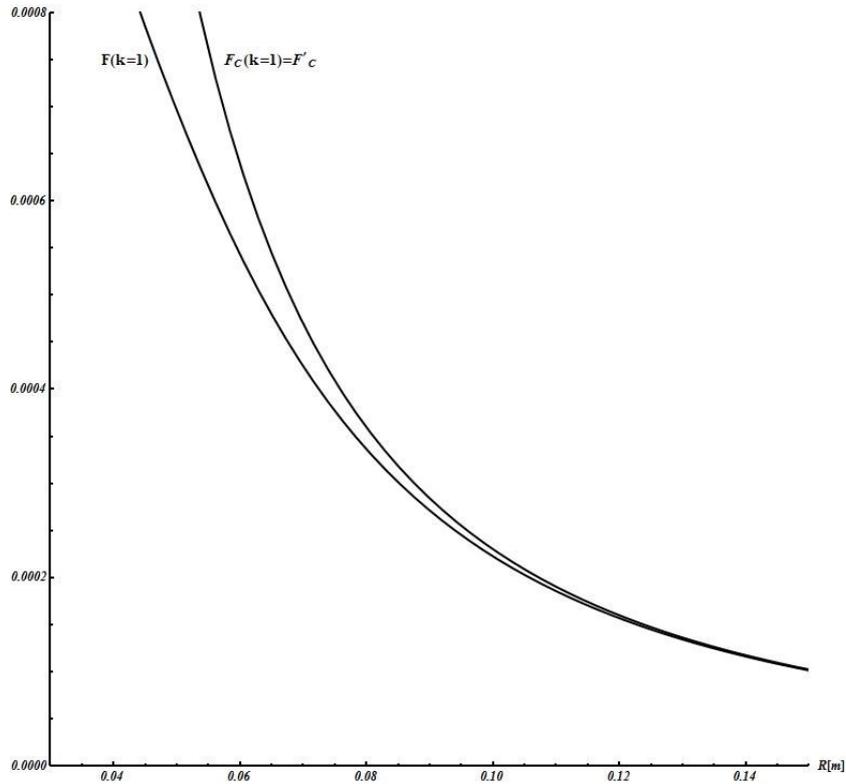

Fig. 6. Graphics of $F_C$ and $F$ with $r_1 = r_2 = 2\times 10^{-2}$ m and $Q_1 = Q_2 = 16\times 10^{-9}$ C for $R \in (4,5]\times 10^{-2}$ m

We can see, that with $k = 1$ is fulfilled $F_C > F > 0$ but the values of $F_C$ and $F$ do not differ significantly, like we mentioned in the third subcase of (*ii*). Therefore, with certain accuracy with $Q_1 = Q_2$ we can assume that $F_C \approx F$.

## 5. Discussion
The Coulomb interactions are fundamental in physics. Thus, it is especially important to evaluate the deviation of these idealized interactions from the real electrostatic interactions between conducting bodies. Moreover, this deviation, as determined for spheres in this article, is significant at small distances between such bodies.

Here, we offer an exact method for determination of the deviation of the Coulomb's force $F_C$ from the force of interaction $F$ between two conducting spheres with different radii and arbitrary charges $Q_1$ and $Q_2$. For the purpose of finding $F_C$ we assume, that $Q_1$ and $Q_2$ are located in the centers of the spheres.

With zero $Q_1 \neq Q_2$ precise the statement of Lekner [5], that at very close distances between the surfaces of the spheres, the force $F$ of interaction between the spheres, is always force of attraction. We find the ratio between the radii and the distance between the centers of the two spheres, charged with like charges, at which the force $F < 0$ (when the charges are unlike, always $F < 0$).

We prove in theory, and specify the experimental results for the interaction of the spheres at relatively close and far distances.

In [12] and [13] numerically are studied the electrostatic interactions, respectively between ellipsoids and tori. The results obtained here could also be used, with approximation, for conducting bodies different than spheres, having center of symmetry, as these bodies are approximated to spheres with equivalent surface areas of the discussed bodies.

Particularly interesting are the cases, when $Q_1 \neq 0$, $Q_2 = 0$ and $Q_1 = Q_2 \neq 0$. In the first case we have established that the force $F$ of interaction between the spheres is short-acting, and $|F|$ is significantly larger than the Coulomb's force $F_C = 0$. In the second case, it turns out that $F$ is not significantly different than $F_C$ and in some studies it could be assumed that $F_C \approx F$. Similar conclusions could be obtained also for the bond energy $W$ between spheres, in relation to the bond energy $W_C$ between point charges.

We use these conclusions in [14], [15], [16], [17], as we find the electrostatic interactions between proton–neutron and proton-proton, modeling them as spheres and tori. Using the established experimental data for nucleons, we obtain that the electrostatic forces between proton – neutron are short-acting, and the bond energy is commensurable to the nuclear, while the forces between proton-proton are long-acting and proportionate to the Coulomb's forces.

The results obtained in this article can be used also in intermolecular and other electrostatic interactions.

### 6. Conclusion

The results in this work offer a theoretical evaluation of the deviation of the Coulomb interaction $F_C$ from the real interaction $F$ between two conducting spheres with equal radii $r_1 = r_2 = r$, charged with arbitrary charges $Q_1$ and $Q_2$. Moreover, we calculate $F_C$ assuming that $Q_1$ and $Q_2$ are located in the centers of the spheres.

When $Q_1 \neq 0$ and $Q_2 \neq 0$, we use the deduced by us exact expression of $F$ through $F_C$, which is a generalization of the interaction between point charges, i.e. $r_1 = r_2 = 0$. It is established, that if $R$ is the distance between the centers of the spheres with radii $r \neq 0$, then with $R/2r \to 1$ the deviation of $F_C$ from $F$ increases, and at relatively small distances between the surfaces of the spheres it is significant. In this case the Coulomb interactions cannot be used.

The Coulomb interactions cannot be used also in the case of $Q_1 \neq 0$ and $Q_2 = 0$ we discus here, in which we find that $F$ is short-acting force.

The article also determines the couple of value $\left(\dfrac{R}{2r}, \dfrac{Q_2}{Q_1}\right)$, in which $F_C \approx F$.

If the radii of the spheres are different, then the deviation coefficients will depend on three, and not on two parameters. In this case two of them must be known, in order to analyze the deviation of the forces of interaction $F_C$ from $F$, as well as the bond energies $W_C$ from $W$, depending on the distance between their surfaces of the spheres. This analysis can be done using the general formulas for $F$ and $W$, which we have obtained [9].

**Acknowledgments**